\documentclass[11pt,letterpaper,superscriptaddress,nofootinbib]{revtex4-1}
\usepackage{amsfonts,amsmath,amsthm,amssymb,amscd}
\usepackage{color,graphicx,array,tikz,pgfplots}
\usepackage[T1]{fontenc}
\usepackage[latin9]{inputenc}
\usepackage{braket}
\usepackage{units}
\makeatletter
\theoremstyle{plain}
\newtheorem{thm}{\protect\theoremname}

\usepackage{hyperref}
\date{}
\usepackage{fullpage}

\makeatother

\usepackage[english]{babel}
\providecommand{\theoremname}{Theorem}
\pgfplotsset{compat=1.14}

\begin{document}

\title{Faster quantum algorithm to simulate Fermionic quantum field theory}
\date{March 29, 2018}

\author{Ali Hamed Moosavian}
\affiliation{Joint Center for Quantum Information and Computer Science, University of Maryland}
\author{Stephen Jordan}
\affiliation{Joint Center for Quantum Information and Computer Science, U. Maryland}
\affiliation{Microsoft, Redmond, WA}

\begin{abstract}
In quantum algorithms discovered so far for simulating scattering processes in quantum field theories, state preparation is the slowest step.  We present an algorithm for preparing particle states to use in simulation of Fermionic Quantum Field Theory (QFT) on a quantum computer, which is based on the matrix product state ansatz. We apply this to the massive Gross-Neveu model in one spatial dimension to illustrate the algorithm, but we believe the same algorithm with slight modifications can be used to simulate any one-dimensional massive fermionic QFT. In the case where the number of particle species is 1, our algorithm can prepare particle states using $O\left( \epsilon^{-3.23\ldots}\right)$ gates, which is much faster than previous known results, namely $O\left(\epsilon^{-8-o\left(1\right)}\right)$. Furthermore, unlike previous methods which were based on adiabatic state preparation, the method given here should be able to simulate quantum phases unconnected to the free theory.
\end{abstract}
\maketitle
\section{Introduction}

Here we analyze the simulation of fermionic quantum field theory (QFT) models on a quantum computer. We use massive Gross-Neveu model to illustrate the procedure. However, the procedures are quite general and can be used to simulate other fermionic systems too. The simulation consists of initializing the incoming particle states on our quantum computer, simulating unitary time evolution according to the lattice quantum field theory Hamiltonian, and then measuring appropriate observables to extract scattering cross sections.
Here we focus exclusively on the state preparation step of the algorithm, because it has been the bottleneck of previous quantum algorithms for simulating scattering in quantum field theories~\cite{Jordan2012,Jordan2014,Jordan2011}.

The original Gross-Neveu model is a relativistic and renormalizable
quantum field theory of $N$ species of fermions in $(1+1)$ space-time
dimensions. It is a rich theory and shares many interesting features
with quantum chromodynamics (QCD), \emph{e.g.,}\ asymptotic freedom and dimensional transmutation~\cite{Gross1974a}. 

Our algorithm simulates scattering of fermionic particles in the Gross-Neveu
quantum field theory with a mass term. The mass term ensures that the theory is gapped, \emph{i.e.,}\ that there is a nonzero energy difference between the ground state and first excited state in the infinite volume limit. This allows us to construct the vacuum (ground state) efficiently by classically computing a Matrix Product State (MPS) description of the vacuum state and then compiling that description directly into a quantum circuit for preparing that state. We then use Rabi oscillations to efficiently excite single-particle wave packets. This completes the state preparation phase of the simulation, after which the scattering of the particles off each other can be simulated using high order Suzuki-Trotter formulas exactly as in~\cite{Jordan2014} or using more recent results for lattice Hamiltonian simulation as in \cite{Haah2018}. Relative to the previous state of the art~\cite{Jordan2014} our state preparation method has better asymptotic complexity in the limit of high precision and is able to simulate the symmetry-broken phase of the Gross-Neveu model, which
was inaccessible to prior state preparation methods, which
simulated an adiabatic process starting from the free theory.

\section{Overview}

\subsection{Gross Neveu Model}
The original Gross-Neveu theory was a QFT describing fermions in $\left(1+1\right)$
space-time dimensions, introduced by Gross and Neveu in 1974 ~\cite{Gross1974a} as a toy model for QCD. The theory has many interesting
features; \emph{e.g.,} similar to QCD, it has asymptotic freedom~\cite{Gross1974a}.
Here, as in~\cite{Jordan2014}, we consider a variant of the Gross-Neveu model in which the
Lagrangian density includes an explicit mass term. Specifically,
\begin{equation}
\mathcal{L}=\sum_{j=1}^{N} \bar{\psi}_{j}\left(i\partial\!\!\!/-m_{0}\right)\psi^{j}+\frac{g^{2}}{2N}\left( \sum_{j=1}^N \bar{\psi}_{j}\psi^{j}\right)^2,\label{eq:Lorentz}
\end{equation}
where each field $\psi_{j}\left(x\right)$ is a two-component spinor,
$\partial\!\!\!/=\sum_{\mu}\gamma^{\mu}\partial_{\mu}$ where $\gamma_{\mu}$
are the 2D Dirac matrices, $m_{0}$ is the bare mass of the model,
$g$ is the coupling constant, and $\bar{\psi}=\psi^{\dagger}\gamma^{0}$.
Outside of high energy physics, this model also has been used in different
branches of condensed matter physics, such as conducting polymers
and systems of strongly correlated electrons~\cite{Lin1998,Thies2005,Thies2005a}.

One can verify that Eq.\ (\ref{eq:Lorentz}) is invariant under Lorentz transformations.
Assuming $m_{0}>0$, the theory has a gap between vacuum and a single particle state. Lorentz-invariance will guarantee the spectrum to be continuous above the first excitation. However, this symmetry and therefore the continuous spectrum are violated when the theory is discretized. Despite this, as discussed in~\cite{Jordan2014}, one can achieve any desired accuracy by using a sufficiently fine lattice spacing. 

Another effect of discretizing the space-time and putting our system
on a lattice would be doubling of species of Dirac fermions, or the
so-called ``Fermion doubling'' problem~\cite{Nielsen1981}. For theories
with chiral symmetry, such as the massless Gross-Neveu model,
there is no way to keep the action real, local and free on a lattice and
preserve translational invariance without getting the extra
fermions~\cite{Nielsen1981a}. However, the mass term breaks chiral symmetry and we can therefore safely solve the fermion doubling problem in the massive case by adding a term to the Hamiltonian, known as the Wilson term, which decouples the extra fermions by giving them large mass~\cite{Wilson1974}.


\subsection{Performance}
\label{sec:performance}

In order to simulate the scattering process on a digital quantum computer,
we first put our system on a spatial lattice of length $L$ and lattice spacing
$a$ with periodic boundary conditions. We then start the algorithm
by preparing the ground state of the resulting lattice quantum field theory
described by the discretized version of Eq.\ (\ref{eq:Lorentz}). There are efficient classical
algorithms for finding the ground state of one dimensional gapped Hamiltonians
as an MPS. There are rigorous upper bounds for the performance of classical algorithms to find the MPS ~\cite{Huang2014,Arad2016}, which are not necessarily applicable to our case (because the norms of individual local terms in the Hamiltonian grow indefinitely as you shrink the lattice spacing $a$). For the purpose of this paper we use classical numerical heuristics such as Density Matrix Renormalization Group (DMRG), that in practice
run in linear time in the number of sites. Specifically the runtime of DMRG in practice is $O\left(n\chi^{3}\right)$;
where $\chi$ is the bond dimension of the matrix product state~\cite{Schollwock2005,Schollwock2011a,White1992a}. Physical arguments show~\cite{Swingle2013,Okunishi1999} that it should suffice to take bond dimension
\begin{equation}
\label{eq:kdef}
\chi=k e^{S_{1/2}} \ ,
\end{equation}
where $S_{1/2}$ is the entanglement entropy between the two half-spaces of the state being prepared, and errors decrease superpolynomially as we increase $k$ beyond unity. For the Gross-Neveu model we can first consider the non-interacting ($g=0$) case, in which the theory splits into $N$ copies of a Dirac quantum field theory. We thus have for the Gross-Neveu entropy $S^{\mathrm{GN}
}(g)$:
\begin{equation}
S_{1/2}^{\textrm{GN}}(g=0) = N S_{1/2}^{\textrm{Dirac}}.
\end{equation}
We are interested in asymptotically high precision. This is the limit where the correlation length $1/m$ is much longer than the lattice spacing $a$. In this limit we obtain the entropy from the conformal field theory describing the massless fermion in one spatial dimension, namely~\cite{Swingle2013, Okunishi1999, Casini, Ginsparg, Qualls}
\begin{equation}
S_{1/2}^{\textrm{Dirac}} \simeq \frac{c}{6} \log \left( \frac{1}{m a} \right) \quad \quad (m a \ll 1)
\end{equation}
where $c=1$ is the central charge. (In the possibly more familiar case of a line segment rather than a half-space one would have twice the entanglement entropy: $\frac{c}{3} \log \left( \frac{1}{m a} \right)$.) Thus, we have
\begin{eqnarray}
\chi & = & k e^{N S_{1/2}^{\mathrm{Dirac}}} \\
     & = & k \left( \frac{1}{m a} \right)^{N/6}.
\end{eqnarray}
The analysis of \cite{Jordan2014} shows that the discretization errors scale as $\epsilon \sim a$. Thus,
\begin{equation}
\label{eq:chipenultimate}
\chi \sim k \epsilon^{-N/6}.
\end{equation}
The relationship between $k$ and error can be understood using the results of \cite{Swingle2013}. Consider Lorentz-invariant 1+1 dimensional quantum field theory discretized onto a lattice of spacing $a$. Let $\lambda_1 \geq \lambda_2 \geq \lambda_3 \ldots$ be the eigenvalues of the reduced density matrix if we divide the vacuum into two halves. In other words, $\lambda_1,\lambda_2,\ldots$ are the Schmidt coefficients in order of decreasing magnitude if we do a Schmidt decomposition of the vacuum state. Then, as shown in \cite{Swingle2013},
\begin{equation}
\label{eq:fn}
f(\chi) \equiv \sum_{j=\chi+1}^\infty \lambda_j \sim \exp \left[ - \frac{(\ln \chi - S_{1/2})^2}{2S_{1/2}} \right]
\end{equation}
for $\chi$ large (\emph{i.e.}\ $\chi > e^{S_{1/2}}$). The magnitude of the error, as measured by trace distance, that we incur by truncating the Schmidt decomposition to some finite bond dimension $\chi$ is captured by $f(\chi)$. From Eq.\ (\ref{eq:fn}) one sees that $f(\chi) = \epsilon$ is achieved by choosing $\chi = k e^{S_{1/2}}$ with
\begin{eqnarray}
\label{eq:kval}
k & = & \exp \left[ \sqrt{2 S_{1/2} \ln (1/\epsilon)} \right] \\
   & = & \exp \left[ \sqrt{ \frac{N}{3} \ln \left( \frac{1}{m a} \right) \ln \left( \frac{1}{\epsilon} \right)} \right] \\
   & \sim & \epsilon^{-\sqrt{N/3}}.
\end{eqnarray}
where the last line follows from $a \sim \epsilon$. 
Combining Eq.\ (\ref{eq:chipenultimate}) and Eq.\ (\ref{eq:kval}) thus yields
\begin{equation}
\chi = O(\epsilon^{-N/6 - \sqrt{N/3}}).
\end{equation}

The classical pre-computation step using DMRG takes time of order $n \chi^3$, where $n$ is the number of lattice sites. Therefore, the complexity of this step is:
\begin{eqnarray}
C_{\mathrm{DMRG}} & = & O\left( n \epsilon^{-N/2 - \sqrt{3N}} \right)\\
& = & O\left( \epsilon^{-N/2 - 1 - \sqrt{3N}} \right)\ , \label{eq:C(DMRG)}
\end{eqnarray}
where the last line follows from $n = L/a$ and $a \sim \epsilon$. As shown in subsequent sections, the quantum state preparation circuit uses $O(\epsilon^{-3-o(1)})$ gates. Thus, the classical preprocessing step is the dominant cost. In particular, for $N=1$ the cost is $O(\epsilon^{-3.23\ldots})$.

An MPS can be compiled into a quantum algorithm for preparing it using Singular Value Decomposition (SVD). The idea is to bring the MPS into a standard form, where each matrix is an isometry that maps the left virtual index to the right virtual index and the physical index. The corresponding quantum circuit applies these isometries one by one on the qubits, until you have the full MPS state. The overall classical subroutine for SVD on the matrices of our system runs in a polynomial time less than $n\chi^3$. 
Applying these quantum isometries on our system would need at most $O\left(n\chi^2\right)$ gates, therefore, this step is not going to be the leading term in the overall performance of the algorithm \cite{Schon2007,Schon2005}.

After preparing the vacuum, the next step is to excite particle states. By introducing sinusoidal source terms in the Hamiltonian, we induce Rabi oscillations to excite two particle wave packets with desired energy and momenta with success probabilities close to $1$. Specifically in section \ref{sec:Rabi Oscillations}, we show that we can prepare a particle state of the interacting theory in time $\frac{\pi}{\lambda}$, with probability
\begin{equation}
P = 1-O\left( \frac{\lambda}{\delta} - \left( \frac{\lambda}{\omega} \right)^2 \right),\label{eq:P}
\end{equation}
%
where $\lambda$ indicates the strength of the source term, $\delta$ is the minimum detuning to higher excited states, and $\omega$ is the energy of the particle compared to the vacuum. According to Eq.\ (\ref{eq:P}), $\frac{\lambda}{\delta}$ is proportional to $1-P$, the chance of failure. We set this to $\epsilon$, which is sufficient to achieve order $\epsilon$ relative error in estimates of scattering cross sections, as defined below in Eq.\ (\ref{eq:scattering amplitude}).  After the particles are prepared we turn off the source term and let them interact
under the Hamiltonian for a desired time. The simulation of the Hamiltonians
can be done using a high order Suzuki-Trotter expansion.

Detailed analysis \cite{Haah2018, Jordan2014} shows that by doing a high
order Suzuki-Trotter expansion the simulation of our Hamiltonian would
require number of gates scaling as:
\begin{equation}
G=O\left(\left(\frac{TL}{a^{2}}\right)^{1+o\left(1\right)}\epsilon^{-o\left(1\right)}\right)\ ,\label{eq:Suzuki-Trotter}
\end{equation}
where $T$ is the total time we want to simulate and $\epsilon$ represents
the desired precision of the calculated scattering amplitude; \emph{i.e.}\
if the discretized algorithm outputs scattering amplitude $\sigma'$
while the actual cross section amplitude of the scattering is $\sigma$,
then:
\begin{equation}
\left(1-\epsilon\right)\sigma\le\sigma'\le\left(1+\epsilon\right)\sigma\ .\label{eq:scattering amplitude}
\end{equation}
So, if the time required to excite a particle is $\frac{\pi}{\lambda}$,
then the number of gates required to prepare the initial particles
would scale as:
\begin{equation}
\begin{array}{rl}
G_{\mathrm{prep}} & =O\left(\left(\frac{L}{a^{2}\lambda}\right)^{1+o\left(1\right)}\epsilon^{-o\left(1\right)}\right)
 
\end{array}
\end{equation}
For high precision results we need $a$ to be proportional to $\epsilon$,
so the overall cost of state preparation would scale as:
\begin{equation}
G_{\mathrm{prep}}=O\left(\epsilon^{-3-o\left(1\right)}\right)\ .
\end{equation}
This is asymptotically smaller than the complexity of the classical DMRG preprocessing
step. Therefore, the overall performance of this algorithm is limited by the classical part of it, $O\left(\epsilon^{-3.23\ldots}\right)$. This is much better than the previous result in \cite{Jordan2014},
$O\left(\epsilon^{-8-o\left(1\right)}\right)$.
To simulate the interaction, we now turn off the source term in the
Hamiltonian and use a similar Suzuki-Trotter expansion to simulate
it. The performance of this part of the algorithm is also given by
Eq.\ (\ref{eq:Suzuki-Trotter}).
After we simulate the scattering of the particles we can read the
outcome by the phase estimation algorithm, that is explained in \cite{Jordan2014}.

\section{Preparing Matrix Product State of the Ground State of the Interacting
Hamiltonian}

We can use Jordan-Wigner transformation to map our fermionic system
to spins \cite{Jordan1928}. Then we can see if our spin Hamiltonian
is local. We also know that there exists a mass gap for the theory.
If the $\left(1+1\right)$-dimensional Hamiltonian is local and gapped, then
we know the ground state obeys an area law\cite{Hastings2007,VanAcoleyen2013}. This
in turn guarantees existence of a matrix product state representation with
low bond dimension~\cite{Vidal2004}.

To simulate the system on a quantum computer we discretize space onto a one-dimensional lattice of spacing $a$. Including the Wilson term the resulting discretized version of the massive Gross-Neveu model is given by the following Hamiltonian.
\begin{equation}
\label{hdiscrete}
H = H_0 + H_g + H_W \,,
\end{equation}
where
\begin{eqnarray}
H_0 & = & \sum_{x \in \Omega} a \sum_{j=1}^N \sum_{\alpha,\beta\in \{0,1\}}\bar{\psi}_{j,\alpha}(x) 
\left[ -i \gamma_{\alpha\beta}^1 \frac{\psi_{j,\beta}(x + a)
- \psi_{j,\beta}(x-a)}{2a} + m_0 \delta_{\alpha,\beta}\psi_{j,\beta}(x) \right] \label{h0} \,, \\
H_g & = & -\frac{g_0^2}{2} \sum_{x \in \Omega} a \bigg( \sum_{j=1}^N\sum_{\alpha\in \{0,1\}}
\bar{\psi}_{j,\alpha} (x) \psi_{j,\alpha}(x) \bigg)^2 \label{hg} \,, \\
H_W & = & \sum_{x \in \Omega} a \sum_{j=1}^N \sum_{\alpha\in \{0,1\}} \left[ - \frac{r}{2a} 
\bar{\psi}_{j,\alpha}(x) \left( \psi_{j,\alpha}(x+a) - 2 \psi_{j,\alpha}(x) + \psi_{j,\alpha}(x-a) \right) 
\right] \,. 
\label{hw} 
\end{eqnarray}
Here, $H_g$ is the interaction term, and $H_W$ is the Wilson term, $1 \le j \le N$ indicates the Fermion species and $0 < r
\leq 1$ is called the Wilson parameter. $H$ is spatially local in the sense 
that it consists only of single-site and nearest-neighbor terms on the
lattice.


All of these Hamiltonian terms consist of pairs of $\bar{\psi}\psi$ terms, and as shown in Appendix \ref{sec: Appendix A}, the Jordan-Wigner transformation yields local spin terms. Also, the massive Gross-Neveu model is gapped, therefore the ground state of this theory obeys area law \cite{Hastings2007}. Therefore, as discussed in \ref{sec:performance} we can find an MPS representation of the ground state in polynomial runtime. However, if we are willing to repeat the simulation to reduce the statistical uncertainty, finding the MPS representation would be a one-time cost and we don't have to repeat it every time we run the quantum part of the simulation. On the other hand, if we change any of the parameters, \emph{e.g.} even the lattice spacing, we will have to repeat this procedure \cite{Huang2014}.


\section{Exciting the state using Rabi oscillations\label{sec:Rabi Oscillations}}

After preparing the vacuum state using matrix product states, the next step is to prepare initial wave packets. We propose doing this by simulating the application of an oscillating perturbation to the Hamiltonian which is on resonance for the creation of a single particle. We do this in two widely separated locations to prepare a pair of wave packets representing particles of high momentum on a collision course. In broad terms, this is the same procedure proposed in the prior algorithm of~\cite{Jordan2014}. However, our analysis here is nonperturbative and consequently achieves tighter error bounds as a function of the strength of the driving term. We can thus improve the complexity of the algorithm by driving the transition more rapidly while maintaining good upper bounds on error.

In more detail, we propose simulating the dynamics induced by
\begin{equation}
\label{eq:hamiltonian}
H(t) = H_0 + \lambda \cos (\omega t) W
\end{equation}
where
\begin{equation}
W = \int dx \left( f(x) \psi_{i,\alpha}(x) + f^*(x) \psi^\dag_{i,\alpha}(x) \right)
\label{Eq:envelope}\end{equation}
and $H_0$ is the unperturbed Gross-Neveu Hamiltonian. Here, $i$ and $\alpha$ are chosen according to the desired type of particle and $f(x)$ is an envelope function that selects the wave packet's location and momentum. For example, $f(x)$ could be taken to be a Gaussian: $f(x) \propto e^{ipx - x^2/\sigma^2}$. As explained in~\cite{Jordan2014}, choosing an appropriate form of this envelope function $f$ selects wave packet states with the desired momentum.

The driving frequency $\omega$ is taken to be on-resonance with the desired transition. If the wave packet is sufficiently broad spatially, and hence narrow in momentum space, then it has a relatively sharply defined energy of $\sqrt{p^2+m^2}$, which we use as our driving frequency $\omega$. Our choice of $f(x)$ ensures that the matrix elements of $W$ coupling to momenta outside the momentum space support of the wave packet are exponentially suppressed. Thus, the main source of error is the possibility of creating the wrong number of particles. The operator $W$ has zero matrix element to create states with even numbers of particles. Thus, the nearest-to-resonance state that can be excited by $W$ is the state of three particles each with momentum $p/3$. The energy of this state is $3 \sqrt{(p/3)^2+m^2}$, which in the ultrarelativistic limit ($p \gg m$) exceeds the energy of the desired state ($\sqrt{p^2+m^2}$) by $~4m^2/p$. The theory of Rabi oscillations shows that if we choose the strength of our driving term $\lambda$ sufficiently small compared to the strength of this detuning $\delta = 4m^2/p$ the probability of exciting this three-particle state can be arbitrarily suppressed. In the following subsections,\ref{subsec:2-level-Approximation} and \ref{subsec:Floquet-Theorem-application}, we make this quantitative. The end result is that the success probability obeys
\begin{equation}
P = 1 - O\left( \lambda/\delta + \lambda^2 t/\delta + \lambda^2/\omega^2 \right)\label{eq:PWithTime}.
\end{equation}
The Rabi rotation from the vacuum to the desired one-particle state takes time $t=\pi/\lambda$. Thus,
\begin{equation}
P = 1 - O\left( \lambda/\delta + \lambda^2/\omega^2 \right).
\end{equation}
To achieve success probability $P = 1-\epsilon$ by simulating this state preparation process using high order Suzuki-Trotter formulae yields a total complexity of $O(\epsilon^{-3-o(1)})$ as discussed in section \ref{sec:performance}.

In the next two subsections,\ref{subsec:2-level-Approximation}
and \ref{subsec:Floquet-Theorem-application} we bound the error we
make when we prepare our particles with this algorithm, and show that
we can prepare particles with probabilities close to $1$. There are
two ingredients to this proof, first in \ref{subsec:2-level-Approximation}
we bound the error we make when we assume our system to be a 2-level
system. Then in subsection \ref{subsec:Floquet-Theorem-application}, we
analyze the resulting 2-level system using Floquet's Theorem to calculate
the probability of successfully exciting one particle.

As explained in~\cite{Jordan2014}, there are two type of failures for the state preparation algorithm. If either or both of the incoming particles are not created, this can be detected by the final measurements of the simulation. The probabilities of exciting zero or more than one particles are suppressed and the following theorems put bounds on them. The latter contributes to the overall precision of the algorithm, $\epsilon$, and it can be controlled by tuning $\lambda$.

We note that our analysis is not fully rigorous in the following sense. To avoid extra technical complexity in sections \ref{subsec:2-level-Approximation} and \ref{subsec:Floquet-Theorem-application} we make two simplifying assumptions. First, we consider the creation of eigenstates rather than wave packets. This can be justified by choosing wave packets whose energy uncertainty is small compared to the inverse duration of the excitation process. Second, in section \ref{subsec:Floquet-Theorem-application} we make the simplifying assumption that the energy levels that we drive a transition between are each nondegenerate and furthermore that the driving operator is purely off-diagonal. In actuality, the excited state we are targeting at momentum $p$ is degenerate with a momentum $-p$ state. However, our choice of envelope function $f$ ensures that matrix elements of the perturbation which couple to the momentum $-p$ state can be exponentially suppressed. Thus, we believe the additional complications associated with a fully realistic treatment do not change the overall scaling. Granting the simplifying assumptions as given, our analysis of the resulting simplified model is fully rigorous.

\subsection{\label{subsec:2-level-Approximation}Few-level Approximation}

In order to estimate and bound the probability to successfully excite a single particle state, we use two approximations. First in this section we look into what is known as the few-level or 2-level approximation and bound the error occurring from this approximation.
In the next subsection we look into the rotating wave approximation
by using an analytical approximate solution to the 2-level driven
system. To avoid inconvenience associated with tracking irrelevant global phases we work with density matrices rather than state vectors. However, the dynamics is closed-system and all states remain pure.
\begin{thm}\label{Theorem 1}
Assume we have a Hamiltonian $H_{0} +\lambda\cos\left(\omega t\right)W$ with $\|W\| = 1$.
Let $\rho(t)$ be the projector onto the state obtained at time $t$ starting with the ground state of $H_0$ at $t=0$. Let $\rho^{(-)}$ be the projector onto the state at time $t$ obtained by projecting $H_{0} +\lambda\cos\left(\omega t\right)W$ onto the span of the lowest $\nu$ eigenstates of $H_0$ and solving the resulting Schrodinger equation in this $\nu$-dimensional Hilbert space. Then:
\begin{equation}
\left|\mathbf{Tr}\left[\rho^{\left(-\right)}\left(t\right)\rho\left(t\right)\right]\right|\ge1-\left(2\nu\lambda+3\nu\lambda^{2}t\right)\frac{1}{\delta}\ , \label{eq: Thm 1}
\end{equation}
where $\delta$ is the minimum detuning from one of the other excited states (\emph{i.e.}\ minimum energy distance between $\{E_0 + \omega, \ldots, E_{\nu-1} + \omega\}$ and the rest of the spectrum of $H_0$). 
\end{thm}
\
From Eq.\ (\ref{eq: Thm 1}) it is evident that smaller values of $\nu$ and larger values of $\delta$ mean a better approximation. For the purpose of this paper $\nu$ is degeneracy of the ground state in the rotating frame, or in other words the number of states that are on-resonance with it.  For an anharmonic system with mirror symmetry between right-going and left-going particles we would have $\nu = 3$. However, with our choice of the envelope function in Eq.\ (\ref{Eq:envelope}), we can effectively break this reflection symmetry of the Hamiltonian and get $\nu=2$. A proof for this theorem is given in Appendix \ref{Appendix B}.

\subsection{\label{subsec:Floquet-Theorem-application}Floquet's Theorem applied to Rabi Oscillations}

In the last subsection we bounded the error incurred by treating the system as a 2-level system. Within the 2-level approximation we analyze the dynamics using Floquet theory, following the treatment of Deng \emph{et al.} \cite{Deng2016}. We thus obtain an upper bound on the probability of remaining in the ground state of the 2-level system. By adding these two sources of error we obtain a bound on total error.


The main ingredient in Deng \emph{et al.}\ analysis is
Floquet's Theorem, which states that if you have a periodic
Hamiltonian in time, there exists a simple change of basis so the
eigenstates have the same periodicity \cite{Floquet1883}.
Floquet's theorem is well studied and is used in a variety of contexts. For example see~\cite{Blekher1992,Cervero1999,Micheli2007} and references therein. Using Floquet's theorem we prove the following error bound.
\begin{thm}\label{Theorem 2}
Assume we have a 2-level Hamiltonian:
\begin{equation}
H=-\frac{\Delta}{2}\mathbb{Z}+\lambda\cos(\omega t)\mathbb{X}\ ,\label{eq:nonrotated hamiltonian}
\end{equation}
where $\mathbb{X}$ and $\mathbb{Z}$ are Pauli matrices and $\Delta>0$ is a real number. We initialize our state at time $t=0$ to be $|0\rangle$ in the standard basis. For $\lambda\ll\omega$, after time $t=\frac{\pi}{\lambda}$,
\begin{equation}
\left|\Braket{ 1|\psi\left(\frac{\pi}{\lambda}\right)} \right|\ge1-\frac{1}{\sqrt{3}}\left(\frac{\lambda}{\omega}\right)^{2}-O\left(\left(\frac{\lambda}{\omega}\right)^{4}\right)\ .
\end{equation}
\end{thm}

A proof for this theorem is presented in Appendix \ref{Appendix C}. 

\subsection{\label{subsec:Total error}Total Error}

Theorem \ref{Theorem 1} shows that the squared inner product between the exact state produced and the state calculated in the 2-level approximation is at least $1-2 \epsilon_1$, where $2 \epsilon_1 = 4 \lambda + 6 \lambda^2 t/\delta$. Hence, up to higher order corrections, the inner product between the exact state produced and the state calculated in the 2-level approximation is $1-\epsilon_1$. Theorem \ref{Theorem 2} shows that the inner product between the state predicted within the 2-level approximation and the desired final state is at least $1-\epsilon_2$, where (neglecting higher order corrections) $\epsilon_2 = \frac{1}{\sqrt{3}} \lambda^2/\omega^2$. Therefore, the inner product between the exact state produced and the desired state is at least $1-\epsilon_1 - \epsilon_2 - 2 \sqrt{\epsilon_1 \epsilon_2}$. We can simplify this expression by noting that $\sqrt{\epsilon_1 \epsilon_2} \leq \max \{\epsilon_1, \epsilon_2\}$. So, the inner product between the exact state obtained and the desired state is $1-O(\epsilon_1 + \epsilon_2)$, and the probability of success, which is the square of this inner product is also $1-O(\epsilon_1 + \epsilon_2)$. This yields Eq.\ (\ref{eq:PWithTime}) and Eq.\  and Eq.\ (\ref{eq:P}).

\section{Energy and Momentum}

After preparing the vacuum of the interacting theory using MPS, we
follow a procedure explained in \cite{Jordan2014}; only a short
summary of this procedure is included here. By using sinusoidal source
terms in the Hamiltonian, we drive our system to excite a specific
energy and momentum. Non-momentum preserving excitations can be neglected
because our choice of envelope function $f$ ensures that they have
exponentially suppressed matrix elements \cite{Jordan2014}.

The renormalized mass $m$ is no longer equal to the bare mass $m_{0}$
in the interacting theory; it's a nontrivial function of bare mass
$m_{0}$, interaction strength $g_{0}$ and lattice spacing $a$. To know what frequency, $\omega$, to use to excite particles it is necessary to know the renormalized mass $m$ . Despite its nontrivial form, in many instances it can be calculated using perturbation theory (as has been done in \cite{Jordan2011,Jordan2014}).
The other approach would be to try and excite a single particle using
guess values of $m$ in the same order of $m_{0}$; then one can use
phase estimation algorithms to check whether they succeeded in exciting
a particle and what is the actual mass of it. For a specific
setup of initial conditions and lattice spacing this would be a one-time
cost. In particular, if one intends to run the algorithm many times for statistical
precision, this cost is incurred only once.

\section{Conclusion}

By introducing our state preparation algorithm, we have significantly improved the performance of quantum algorithms for simulation of Fermionic QFT. Furthermore, unlike adiabatic state preparation, MPS-based state preparation should be applicable to the phases of the theory which are unconnected to the free theory. In particular, starting from the free theory, as one increases the coupling constant while keeping the bare mass fixed, the Gross-Neveu model exhibits a quantum phase transition at which the eigenvalue gap (\emph{i.e.}\ physical mass) vanishes. On the other side of this transition the $\psi \to -\psi$ symmetry is spontaneously broken. Because of the vanishing gap (at least in the infinite volume limit) adiabatic state preparation may have problems producing the vacuum of the symmetry-broken phase. However, an MPS-based method should be able to access this phase directly without having to cross a phase transition.

Although we only used Gross-Neveu to illustrate our new state preparation method, it should be applicable to other Fermionic models in one spatial dimension. Extending the techniques to two spatial dimensions, such as through the uses of Projected Entangled Pair States (PEPS)~\cite{Schwarz2011,Schwarz2012} is an interesting avenue for future research.

\section{Acknowledgements}

We thank Norbert Schuch, Brian Swingle, and Scott Glancy for helpful discussions. This work was supported by a NIST Building The Future grant and by Department of Energy grant DE-SC0016431. Parts of this manuscript were completed while SJ was an employee of NIST. These parts are thus a contribution of NIST, an agency of the US government, and are therefore not subject to US copyright.
\newpage
\appendix
\section{Locality Of The Equivalent Spin Hamiltonian\label{sec: Appendix A}}
Here we will use Jordan-Wigner transformation and derive the mapped Hamiltonian terms explicitly. Fermionic systems obey anti-commutation relations that cause the states to be non-local. 
\begin{equation}
\left\{ \psi_{j,\alpha}\left(x\right),\psi_{k,\beta}^{\dagger}\left(y\right)\right\} = a^{-1} \delta_{j,k}\delta_{\alpha,\beta}\delta_{x,y} \mathbb{I} ,
\end{equation}
\begin{equation}
\left\{ \psi_{j,\alpha}^{\dagger}\left(x\right),\psi_{k,\beta}^{\dagger}\left(y\right)\right\} = \left\{ \psi_{j,\alpha}\left(x\right),\psi_{k,\beta}\left(y\right)\right\} = 0 ,
\end{equation}
where $\delta_{m,n}$ is the Kronecker Delta function, $j$ and $k$ represent different Fermion species and $\alpha$ and $\beta$ indicate matter and antimatter particles. The Jordan-Wigner transformation is defined as:{\small{}
\begin{align}
\psi_{j,0}\left(\eta a\right) & \to \frac{-1}{\sqrt{a}}\bigotimes_{\kappa=1}^{\eta-1}\left(\bigotimes_{\xi=1}^{N}\left(\mathbb{Z}_{\left(j,0\right)}^{\left(\kappa\right)}\otimes \mathbb{Z}_{\left(j,1\right)}^{\left(\kappa\right)}\right)\right)\otimes \bigotimes_{\xi=1}^{j-1}\left(\mathbb{Z}_{\left(j,0\right)}^{\left(\eta\right)}\otimes \mathbb{Z}_{\left(j,1\right)}^{\left(\eta\right)}\right) \otimes  \left(a^+\right)_{\left(j,0\right)}^{\left(\eta\right)}\otimes \mathbb{I}_{\left(j,1\right)}^{\left(\eta\right)}\ , \\
\psi_{j,1}\left(\eta a\right) & \to \frac{-1}{\sqrt{a}}\bigotimes_{\kappa=1}^{\eta-1}\left(\bigotimes_{\xi=1}^{N}\left(\mathbb{Z}_{\left(j,0\right)}^{\left(\kappa\right)}\otimes \mathbb{Z}_{\left(j,1\right)}^{\left(\kappa\right)}\right)\right)\otimes \bigotimes_{\xi=1}^{j-1}\left(\mathbb{Z}_{\left(j,0\right)}^{\left(\eta\right)}\otimes \mathbb{Z}_{\left(j,1\right)}^{\left(\eta\right)}\right) \otimes \mathbb{Z}_{\left(j,0\right)}^{\left(\eta\right)} \otimes \left(a^+\right)_{\left(j,1\right)}^{\left(\eta\right)} \ , \\
\Rightarrow\psi_{j,0}^{\dagger}\left(\eta a\right) & \to \frac{-1}{\sqrt{a}}\bigotimes_{\kappa=1}^{\eta-1}\left(\bigotimes_{\xi=1}^{N}\left(\mathbb{Z}_{\left(j,0\right)}^{\left(\kappa\right)}\otimes \mathbb{Z}_{\left(j,1\right)}^{\left(\kappa\right)}\right)\right)\otimes \bigotimes_{\xi=1}^{j-1}\left(\mathbb{Z}_{\left(j,0\right)}^{\left(\eta\right)}\otimes \mathbb{Z}_{\left(j,1\right)}^{\left(\eta\right)}\right) \otimes  \left(a^-\right)_{\left(j,0\right)}^{\left(\eta\right)}\otimes \mathbb{I}_{\left(j,1\right)}^{\left(\eta\right)}\ , \\
\Rightarrow\psi_{j,1}^{\dagger}\left(\eta a\right) & \to \frac{-1}{\sqrt{a}}\bigotimes_{\kappa=1}^{\eta-1}\left(\bigotimes_{\xi=1}^{N}\left(\mathbb{Z}_{\left(j,0\right)}^{\left(\kappa\right)}\otimes \mathbb{Z}_{\left(j,1\right)}^{\left(\kappa\right)}\right)\right)\otimes \bigotimes_{\xi=1}^{j-1}\left(\mathbb{Z}_{\left(j,0\right)}^{\left(\eta\right)}\otimes \mathbb{Z}_{\left(j,1\right)}^{\left(\eta\right)}\right)  \otimes \mathbb{Z}_{\left(j,0\right)}^{\left(\eta\right)} \otimes \left(a^-\right)_{\left(j,1\right)}^{\left(\eta\right)} \ , \\
\Rightarrow\bar{\psi}_{j,0}\left(\eta a\right) & \to \frac{-i}{\sqrt{a}}\bigotimes_{\kappa=1}^{\eta-1}\left(\bigotimes_{\xi=1}^{N}\left(\mathbb{Z}_{\left(j,0\right)}^{\left(\kappa\right)}\otimes \mathbb{Z}_{\left(j,1\right)}^{\left(\kappa\right)}\right)\right)\otimes \bigotimes_{\xi=1}^{j-1}\left(\mathbb{Z}_{\left(j,0\right)}^{\left(\eta\right)}\otimes \mathbb{Z}_{\left(j,1\right)}^{\left(\eta\right)}\right) \otimes \mathbb{Z}_{\left(j,0\right)}^{\left(\eta\right)} \otimes \left(a^-\right)_{\left(j,1\right)}^{\left(\eta\right)} \ , \\
\Rightarrow\bar{\psi}_{j,1}\left(\eta a\right) & \to \frac{i}{\sqrt{a}}\bigotimes_{\kappa=1}^{\eta-1}\left(\bigotimes_{\xi=1}^{N}\left(\mathbb{Z}_{\left(j,0\right)}^{\left(\kappa\right)}\otimes \mathbb{Z}_{\left(j,1\right)}^{\left(\kappa\right)}\right)\right)\otimes \bigotimes_{\xi=1}^{j-1}\left(\mathbb{Z}_{\left(j,0\right)}^{\left(\eta\right)}\otimes \mathbb{Z}_{\left(j,1\right)}^{\left(\eta\right)}\right)  \otimes  \left(a^-\right)_{\left(j,0\right)}^{\left(\eta\right)}\otimes \mathbb{I}_{\left(j,1\right)}^{\left(\eta\right)}\ ,
\end{align}}
where $a+=\left( \begin{array}{cc}
0 & 0\\
1 & 0
\end{array}\right)$ and $a^- =\left( \begin{array}{cc}
0 & 1\\
0 & 0
\end{array}\right)$.
As one can easily verify, the long tails of $\mathbb{Z}$ tensor products cancel out and we are left with local Hamiltonian terms. The Hamiltonian terms are explicitly mapped to:{\small{}
\begin{align}
H_{0} &\to  \sum_{j=1}^{N}\sum_{\eta a\in\Omega}\left[\frac{i}{2a}\left(\left(a^-\right)_{\left(j,1\right)}^{\left(\eta\right)}\otimes \mathbb{Z}^{\otimes 2N -1} \otimes \left(a^+\right)_{\left(j,1\right)}^{\left(\eta+1\right)}-\left(a^-\right)_{(j,0)}^{\left(\eta\right)}\otimes\mathbb{Z}^{\otimes 2N-1}\otimes\left(a^+\right)_{(j,0)}^{\left(\eta+1\right)}\right)\right.\nonumber \\
 & -  \left. i m_{0}\left(a^+\right)_{(j,0)}^{\left(\eta\right)}\otimes\left(a^+\right)_{(j,1)}^{\left(\eta\right)}+h.c.\right]\ , \\
H_{g} & \to  \frac{g_{0}^{2}}{2a}\sum_{\eta a\in\Omega}\left[\sum_{j=1}^{N}\left(\left(a^-\right)_{(j,0)}^{\left(\eta\right)}\otimes\left(a^+\right)_{(j,1)}^{\left(\eta\right)}-\left(a^+\right)_{(j,0)}^{\left(\eta\right)}\otimes\left(a^-\right)_{(j,1)}^{\left(\eta\right)}\right)\right]^{2}\ , \\
H_{w} & \to  \sum_{\eta a\in\Omega}\frac{r}{2a}\sum_{j=1}^{N}\left[i\left(a^-\right)_{(j,1)}^{\left(\eta\right)}\otimes \mathbb{Z}^{\otimes 2N-2}\otimes\left(a^+\right)_{(j,0)}^{\left(\eta+1\right)}+i\left(a^+\right)_{(j,0)}^{\left(\eta\right)}\otimes \mathbb{Z}^{\otimes 2N}\otimes\left(a^i\right)_{(j,1)}^{\left(\eta+1\right)}\right. \nonumber\\
& -  \left.2i \left(a^+\right)_{(j,0)}^{\left(\eta\right)}\otimes\left(a^-\right)_{(j,1)}^{\left(\eta\right)}+h.c.\right]\ .
\end{align}
}And as we were expecting, all of these Hamiltonian terms are indeed local.

\section{Bounds On The Error Incurred From Few-level Approximation \label{Appendix B}}

Here we present the proof of Theorem \ref{Theorem 1}:
\begin{proof}
Starting with the same Hamiltonian as Eq.\ (\ref{eq:hamiltonian}) we
have:

\begin{equation}
H=H_{0}+\lambda\cos\left(\omega t\right)W\ ,
\end{equation}
\begin{equation}
H_{0}|\psi_{j}\rangle=E_{j}|\psi_{j}\rangle\ .
\end{equation}
Let's define $\omega_{ij}$ as:
\begin{equation}
\omega_{ij}=E_{i}-E_{j}\ .
\end{equation}
Now, let's go to the interaction picture. That is, for any operator $O$ let $O_I = e^{i H_0 t} O e^{-i H_0 t}$. In particular, we will solve for the dynamics of 
\begin{equation}
\rho_I = e^{i H_0 t} \rho e^{-i H_0 t}.
\end{equation}
The interaction picture is convenient for treating the time-dependent perturbation terms in the Hamiltonian. For the evolution equation we  have:
\begin{equation}
i\frac{d}{dt}\rho_{I}=\lambda\cos\left(\omega t\right)[W_{I},\rho_{I}]\ ,
\end{equation}
where: 
\begin{equation}
W_{I}\left(t\right)=e^{iH_{0}t}We^{-iH_{0}t}\ .\label{eq:W,Interaction picture}
\end{equation}
Note that our states are pure and we can switch between density matrix
and state representation for convenience. Now let's decompose the
Hilbert space into $\mathcal{H=}\mathcal{H}^{\left(+\right)}\oplus\mathcal{H}^{\left(-\right)}$
where:
\begin{equation}
\mathcal{H}^{\left(-\right)}=\text{span}\left\{ |\psi_{0}\rangle,|\psi_{1}\rangle,\cdots,|\psi_{\nu-1}\rangle\right\} \ ,
\end{equation}
\begin{equation}
\mathcal{H}^{\left(+\right)}=\text{span}\left\{ |\psi_{\nu}\rangle,|\psi_{\nu+1}\rangle,\cdots\right\} \ .
\end{equation}
We are going to consider the more general few-level approximation;
the few-level approximation is more applicable than the 2-level approximation when one's dealing with a system which has degeneracies. In this section we are trying
to bound the error that incurs by limiting our Hilbert space to $\mathcal{H}^{\left(-\right)}$.

Let $P^{\left(A\right)}$ be the projector onto $\mathcal{H}^{\left(A\right)}$
for $A\in\left\{ +,-\right\} $ and 
\begin{equation}
W_{I}^{\left(AB\right)}=P^{\left(A\right)}W_{I}P^{\left(B\right)}\ .
\end{equation}
The initial state is the ground state of the theory:
\begin{equation}
\rho_{I}\left(0\right)=\left|\psi_{0}\left\rangle \right\langle \psi_{0}\right|\ .
\end{equation}

The few-level approximation is defined as:
\begin{equation}
\rho_{I}\left(t\right)\simeq\rho_{I}^{\left(-\right)}\left(t\right)\ ,
\end{equation}
where $\rho_{I}^{\left(-\right)}\left(t\right)$ is defined by:
\begin{equation}
\rho_{I}^{\left(-\right)}\left(0\right)=\rho_{I}\left(0\right)=\left|\psi_{0}\left\rangle \right\langle \psi_{0}\right|\ ,
\end{equation}
\begin{equation}
i\frac{d}{dt}\rho_{I}^{\left(-\right)}=\lambda\cos\left(\omega t\right)[W_{I}^{\left(--\right)},\rho_{I}^{\left(-\right)}]\ .\label{eq:d/dt psi_i}
\end{equation}
Note that $\rho_{I}^{\left(-\right)}\ne P^{\left(-\right)}\rho_{I}P^{\left(-\right)}$.
For simplicity, let's restrict our attention to the case that $W$
has no diagonal terms, \emph{i.e.}\
\begin{equation}
\left\langle \psi_{j}\left|W\right|\psi_{j}\right\rangle =0\text{ , }\forall j\in\mathbb{N\ }.
\end{equation}
Furthermore we assume the system is on resonance, $\omega_{10}=\omega.$
We use the trace, $\mathbf{Tr}\left[\rho_{I}^{\left(-\right)}\left(t\right)\rho_{I}\left(t\right)\right]$,
to quantify the error. We have:
\begin{equation}
\begin{array}{rl}
i\frac{d}{dt}\mathbf{Tr}\left[\rho_{I}^{\left(-\right)}\left(t\right)\rho_{I}\left(t\right)\right] & =i\mathbf{Tr}\left[\frac{d}{dt}\left(\rho_{I}^{\left(-\right)}\left(t\right)\rho_{I}\left(t\right)\right)\right] \vspace{3pt} \\
 & =\mathbf{Tr}\left[\lambda\cos\left(\omega t\right)\left[W_{I}^{\left(--\right)},\rho_{I}^{\left(-\right)}\right]\rho_{I}+\lambda\cos\left(\omega t\right)\rho_{I}^{\left(-\right)}\left[W_{I},\rho_{I}\right]\right] \vspace{3pt}\\
 & =\lambda\cos\left(\omega t\right)\mathbf{Tr}\left[\left(W_{I}^{\left(--\right)}-W_{I}\right)\left[\rho_{I}^{\left(-\right)},\rho_{I}\right]\right] \vspace{3pt}\\
 & =\lambda\cos\left(\omega t\right)\mathbf{Tr}\left[W_{I}^{\left(-+\right)}\left(t\right)\rho_{I}\left(t\right)\rho_{I}^{\left(-\right)}\left(t\right)-W_{I}^{\left(+-\right)}\left(t\right)\rho_{I}^{\left(-\right)}\left(t\right)\rho_{I}\left(t\right)\right]\ .
\end{array}
\end{equation}
The integral form of this equation becomes:
\begin{equation}
\mathbf{Tr}\left[\rho_{I}^{\left(-\right)}\left(t\right)\rho_{I}\left(t\right)\right]=1-i\lambda\int_{0}^{t}d\tau\cos\left(\omega t\right)\mathbf{Tr}\left[W_{I}^{\left(-+\right)}\left(t\right)\rho_{I}\left(t\right)\rho_{I}^{\left(-\right)}\left(t\right)-W_{I}^{\left(+-\right)}\left(t\right)\rho_{I}^{\left(-\right)}\left(t\right)\rho_{I}\left(t\right)\right]\ .
\end{equation}
Now, if we expand the cosine function as the sum of two exponentials,
we can write:
\begin{equation}
\left|\mathbf{Tr}\left[\rho_{I}^{\left(-\right)}\left(t\right)\rho_{I}\left(t\right)\right]\right|\ge1-K_{1}-K_{2}-L_{1}-L_{2}\ ,\label{eq:Tr rho-rho}
\end{equation}
where
\begin{equation}
K_{k}=\frac{\lambda}{2}\left|\int_{0}^{t}d\tau e^{\left(-1\right)^{k}i\omega\tau}\mathbf{Tr}\left[W_{I}^{\left(-+\right)}\left(t\right)\rho_{I}\left(t\right)\rho_{I}^{\left(-\right)}\left(t\right)\right]\right|\ ,
\end{equation}
\begin{equation}
L_{k}=\frac{\lambda}{2}\left|\int_{0}^{t}d\tau e^{\left(-1\right)^{k}i\omega\tau}\mathbf{Tr}\left[W_{I}^{\left(+-\right)}\left(t\right)\rho_{I}^{\left(-\right)}\left(t\right)\rho_{I}\left(t\right)\right]\right|\ .
\end{equation}
Using Eq.\ (\ref{eq:W,Interaction picture}) we can rewrite $K_{k}$
as:
\begin{equation}
\begin{array}{rl}
K_{k} & =\frac{\lambda}{2}\left|\int_{0}^{t}d\tau e^{\left(-1\right)^{k}i\omega\tau}\sum_{\mu,\eta}\left\langle \psi_{I}^{\left(-\right)}\left(\tau\right)\left|_{\mu}W_{I}^{\left(-+\right)}\left(\tau\right)_{\mu\eta}\right|\psi_{I}\left(\tau\right)\right\rangle _{\eta}\Braket{ \psi_{I}\left(\tau\right)|\psi_{I}^{\left(-\right)}\left(\tau\right)} \right| \vspace{4pt}\\
 & =\frac{\lambda}{2}\left|\int_{0}^{t}d\tau e^{\left(-1\right)^{k}i\omega\tau}\sum_{\mu=0}^{\nu-1}\sum_{\eta\ge m}\left\langle \psi_{I}^{\left(-\right)}\left(\tau\right)\left|_{\mu}\hat{W}_{\mu\eta}e^{i\omega_{\mu\eta}\tau}\right|\psi_{I}\left(\tau\right)\right\rangle _{\eta}\Braket{ \psi_{I}\left(\tau\right)|\psi_{I}^{\left(-\right)}\left(\tau\right)} \right| \vspace{4pt}\\
 & =\frac{\lambda}{2}\left|\sum_{\mu=0}^{\nu-1}\sum_{\eta\ge \nu}\int_{0}^{t}d\tau e^{\left(-1\right)^{k}i\omega\tau+i\omega_{\mu\eta}\tau}\left\langle \psi_{I}^{\left(-\right)}\left(\tau\right)\left|_{\mu}\hat{W}_{\mu\eta}\right|\psi_{I}\left(\tau\right)\right\rangle _{\eta}\Braket{ \psi_{I}\left(\tau\right)|\psi_{I}^{\left(-\right)}\left(\tau\right)} \right|\ .
\end{array}\label{eq:K_k}
\end{equation}
We calculate this integral using integration by parts:
\begin{equation}
\begin{array}{rl}
I_{k\mu\eta} & \equiv\int_{0}^{t}d\tau e^{\left(-1\right)^{k}i\omega\tau+i\omega_{\mu\eta}\tau}\left\langle \psi_{I}^{\left(-\right)}\left(\tau\right)\left|_{\mu}\hat{W}_{\mu\eta}\right|\psi_{I}\left(\tau\right)\right\rangle _{\eta}\Braket{ \psi_{I}\left(\tau\right)|\psi_{I}^{\left(-\right)}\left(\tau\right)} \ ,\\
du_{k\mu\eta} & \equiv d\tau e^{\left(-1\right)^{k}i\omega\tau+i\omega_{\mu\eta}\tau}\ ,\\
v_{\mu\eta} & \equiv\left\langle \psi_{I}^{\left(-\right)}\left(\tau\right)\left|_{\mu}\hat{W}_{\mu\eta}\right|\psi_{I}\left(\tau\right)\right\rangle _{\eta}\Braket{ \psi_{I}\left(\tau\right)|\psi_{I}^{\left(-\right)}\left(\tau\right)} \ .
\end{array}
\end{equation}
\begin{equation}
\Rightarrow\left\{ \begin{array}{rl}
u_{k\mu\eta} & =\frac{e^{\left[\left(-1\right)^{k}\omega+\omega_{\mu\eta}\right]i\tau}}{i\left[\left(-1\right)^{k}\omega+\omega_{\mu\eta}\right]}\\
dv_{\mu\eta} & =d\tau\frac{\lambda\cos\left(\omega\tau\right)}{i}\mathbf{Tr}\left[-W_{I}^{\left(+-\right)}\rho_{I}^{\left(-\right)}\hat{W}_{\mu\eta}\rho_{I}+\left(\hat{W}_{\mu\eta}\left(W_{I}^{\left(++\right)}+W_{I}^{\left(+-\right)}\right)-W_{I}^{\left(--\right)}\hat{W}_{\mu\eta}\right)\rho_{I}\rho_{I}^{\left(-\right)}\right]
\end{array}\right.\ .
\end{equation}
So $I_{k\mu\eta}$ becomes:
\begin{equation}
\begin{array}{rl}
I_{k\mu\eta} & =\left.\frac{1}{i}\mathbf{Tr}\left[\rho_{I}^{\left(-\right)}\left(t\right)\hat{R}_{k\mu\eta}\rho_{I}\left(t\right)\right]\right|_{\tau=0}^{t} \vspace{4pt}\\
 & +\int_{0}^{t}d\tau\lambda\cos\left(\omega\tau\right)\mathbf{Tr}\left[-W_{I}^{\left(+-\right)}\rho_{I}^{\left(-\right)}\hat{R}_{k\mu\eta}\rho_{I}+\left(\hat{R}_{k\mu\eta}\left(W_{I}^{\left(++\right)}+W_{I}^{\left(+-\right)}\right)-W_{I}^{\left(--\right)}\hat{R}_{k\mu\eta}\right)\rho_{I}\rho_{I}^{\left(-\right)}\right]\ ,
\end{array}
\end{equation}
where $\hat{R}_{k\mu\eta}\left(\tau\right)$ is defined as:
\begin{equation}
\hat{R}_{k\mu\eta}\left(\tau\right)=\frac{e^{\left[\left(-1\right)^{k}\omega+\omega_{\mu\eta}\right]i\tau}}{\left[\left(-1\right)^{k}\omega+\omega_{\mu\eta}\right]}W_{\mu\eta}\left|\left.\psi_{\mu}\right\rangle \left\langle \psi_{\eta}\right.\right|\ .\label{eq:Rkmueta}
\end{equation}

Now putting this back into Eq.\ (\ref{eq:K_k}) we get:
\begin{equation}
\begin{array}{rl}
K_{k} & =\frac{\lambda}{2}\left|\frac{1}{i}\Braket{ \psi_{I}^{\left(-\right)}\left(t\right)\left|R_{k}\left(t\right)\right|\psi_{I}\left(t\right)} \left\langle \psi_{I}\left(t\right)|\psi_{I}^{\left(-\right)}\right\rangle -\frac{1}{i}\left\langle \psi_{I}\left(0\right)\left|R_{k}\left(0\right)\right|\psi_{I}\left(0\right)\right\rangle \right. \vspace{4pt}\\
 & \left.+\int_{0}^{t}d\tau\lambda\cos\left(\omega\tau\right)\left(\left\langle \psi_{I}\left|-W_{I}^{\left(+-\right)}\rho_{I}^{\left(-\right)}R_{k}+\rho_{I}^{\left(-\right)}\left(R_{k}\left(W_{I}^{\left(++\right)}+W_{I}^{\left(+-\right)}\right)-W_{I}^{\left(--\right)}R_{k}\right)\right|\psi_{I}\right\rangle \right)\right|\ ,
\end{array}
\end{equation}
where from Eq.\ (\ref{eq:Rkmueta}):
\begin{equation}
R_{k}=\sum_{\mu=0}^{\nu-1}\sum_{\eta\ge \nu}\frac{e^{\left[\left(-1\right)^{k}\omega+\omega_{\mu\eta}\right]i\tau}}{\left[\left(-1\right)^{k}\omega+\omega_{\mu\eta}\right]}W_{\mu\eta}\left|\left.\psi_{\mu}\right\rangle \left\langle \psi_{\eta}\right.\ \right|.
\end{equation}
Hence, by the triangle inequality and submultiplicativity of the operator
norm:
\begin{equation}
\begin{array}{rl}
K_{k} & \le\frac{\lambda}{2}\left(\left\Vert R_{k}\left(t\right)\right\Vert +\left\Vert R_{k}\left(0\right)\right\Vert \right) \vspace{4pt}\\
 & +\frac{\lambda^{2}t}{2}\max_{0\le\tau\le t}\left(\left(\left\Vert W_{I}^{\left(++\right)}\left(\tau\right)+W_{I}^{\left(+-\right)}\left(\tau\right)\right\Vert +\left\Vert W_{I}^{\left(-+\right)}\left(\tau\right)\right\Vert +\left\Vert W_{I}^{\left(--\right)}\left(\tau\right)\right\Vert \right)\left\Vert R_{k}\left(\tau\right)\right\Vert \right) \vspace{4pt}\\
 & \le\frac{\lambda}{2}\left(\left\Vert R_{k}\left(t\right)\right\Vert +\left\Vert R_{k}\left(0\right)\right\Vert +3\lambda t\max_{0\le\tau\le t}\left(\left\Vert W_{I}\left(\tau\right)\right\Vert \left\Vert R_{k}\left(\tau\right)\right\Vert \right)\right) \vspace{4pt}\\
 & \le\frac{\lambda}{2}\left(\left\Vert R_{k}\left(t\right)\right\Vert +\left\Vert R_{k}\left(0\right)\right\Vert +3\lambda t\left\Vert W\right\Vert \max_{0\le\tau\le t}\left(\left\Vert R_{k}\left(\tau\right)\right\Vert \right)\right)\ .
\end{array}\label{eq:K_k inequality}
\end{equation}
Note that $\left\Vert W_{I}\left(\tau\right)\right\Vert =\left\Vert W\right\Vert =1$
for all $\tau$. We can decompose $R_{k}\left(\tau\right)$ as the
sum over the first index as:
\begin{equation}
R_{k}\left(\tau\right)=\sum_{\mu=0}^{\nu-1}R_{k}^{\mu}\left(\tau\right)\ ,
\end{equation}
where
\begin{equation}
R_{k}^{\mu}\left(\tau\right)=\sum_{\eta\ge \nu}\frac{e^{\left[\left(-1\right)^{k}\omega+\omega_{\mu\eta}\right]i\tau}}{\left[\left(-1\right)^{k}\omega+\omega_{\mu\eta}\right]}W_{\mu\eta}\left|\left.\psi_{\mu}\right\rangle \left\langle \psi_{\eta}\right.\right|\text{ \ \ \ \ \ \ \ }\mu=0,1,\cdots,\nu-1\ .
\end{equation}
So
\begin{equation}
\left\Vert R_{k}\left(\tau\right)\right\Vert \le\left\Vert R_{k}^{0}\left(\tau\right)\right\Vert +\left\Vert R_{k}^{1}\left(\tau\right)\right\Vert +\cdots+\left\Vert R_{k}^{\nu-1}\left(\tau\right)\right\Vert \label{eq:R_k decomposition}
\end{equation}
and
\begin{equation}
R_{k}^{\mu}\left(\tau\right)=P_{\mu}WD_{k}^{\mu}U_{k}^{\mu}\left(\tau\right)\ ,
\end{equation}
where
\begin{equation}
\begin{array}{rl}
P_{\mu} & =\left|\left.\psi_{\mu}\right\rangle \left\langle \psi_{\mu}\right.\right|\ \ \ \ \ \ \ \mu=0,1,\cdots,\nu-1\ ,\vspace{4pt}\\
D_{k}^{\mu} & =\sum_{\eta\ge \nu}\frac{\left|\left.\psi_{\eta}\right\rangle \left\langle \psi_{\eta}\right.\right|}{\left[\left(-1\right)^{k}\omega+\omega_{\mu\eta}\right]}\ ,\vspace{4pt}\\
U_{k}^{\mu}\left(\tau\right) & =\sum_{\eta=0}^{\infty}e^{\left[\left(-1\right)^{k}\omega+\omega_{\mu\eta}\right]i\tau}\left|\left.\psi_{\eta}\right\rangle \left\langle \psi_{\eta}\right.\right|\ .
\end{array}
\end{equation}
One sees that $U_{k}^{\mu}\left(\tau\right)$ is unitary. So
\begin{equation}
\left\Vert R_{k}^{\mu}\left(\tau\right)\right\Vert \le\left\Vert P_{\mu}\right\Vert \cdot \left\Vert W\right\Vert \cdot \left\Vert D_{k}^{\mu}\right\Vert.\label{eq:abs Rkmu}
\end{equation}
$P_{\mu}$ is a projector so $\left\Vert P_{\mu}\right\Vert =1$.
$\left\Vert D_{k}^{\mu}\right\Vert $ is diagonal so it's easy to
read its spectral norm:
\begin{equation}
\left\Vert D_{k}^{\mu}\right\Vert =\frac{1}{\delta_{k}^{\mu}}\ ,
\end{equation}
where 
\begin{equation}
\delta_{k}^{\mu}=\min_{\eta>\nu}\left|\left(-1\right)^{k}\omega+\omega_{\mu\eta}\right|\ .
\end{equation}
Thus Eq.\ (\ref{eq:abs Rkmu}) yields
\begin{equation}
\left\Vert R_{k}^{\mu}\left(\tau\right)\right\Vert \le\frac{1}{\delta_{k}^{\mu}}\ .\label{eq:Rkmu inequality}
\end{equation}
Stitching equations (\ref{eq:Rkmu inequality}), (\ref{eq:R_k decomposition}) and (\ref{eq:K_k inequality}) together one gets:
\begin{equation}
K_{k}\le\frac{\lambda}{2}\sum_{\mu=0}^{\nu-1}\left(\frac{2}{\delta_{k}^{\mu}}+3\lambda t\frac{1}{\delta_{k}^{\mu}}\right)\ .
\end{equation}
Following the same procedure for $L_{k}$, one gets the same bound:
\begin{equation}
L_{k}\le\frac{\lambda}{2}\sum_{\mu=0}^{\nu-1}\left(\frac{2}{\delta_{k}^{\mu}}+3\lambda t\frac{1}{\delta_{k}^{\mu}}\right)\ .
\end{equation}
Hence Eq.\ (\ref{eq:Tr rho-rho}) yields
\begin{equation}
\left|\mathbf{Tr}\left[\rho_{I}^{\left(-\right)}\left(t\right)\rho_{I}\left(t\right)\right]\right|\ge1-\left(\lambda+\frac{3\lambda^{2}t}{2}\right)\sum_{\mu=0}^{\nu-1}\sum_{k=1}^{2}\frac{1}{\delta_{k}^{\mu}}\ .
\end{equation}
One sees that 
\begin{equation}
\sum_{\mu=0}^{\nu-1}\sum_{k=1}^{2}\frac{1}{\delta_{k}^{\mu}}\le\frac{2\nu}{\delta}\ ,
\end{equation}
where 
\begin{equation}
\delta=\min\left\{ \delta_{k}^{\mu}|\mu\in\left\{ 0,1,\cdots,\nu-1\right\},\ k\in\left\{ 1,2\right\} \right\} \ .
\end{equation}
In other words, $\delta$ is the detuning to the nearest off-resonant transition
from the first $\nu$ states. So the error caused by the $\nu$-level
approximation can be bounded:
\begin{equation}
\left|\mathbf{Tr}\left[\rho_{I}^{\left(-\right)}\left(t\right)\rho_{I}\left(t\right)\right]\right|=\left|\mathbf{Tr}\left[\rho^{\left(-\right)}\left(t\right)\rho\left(t\right)\right]\right|\ge1-\left(2\nu\lambda+3\nu\lambda^{2}t\right)\frac{1}{\delta}\ .
\end{equation}
\end{proof}

\section{Bounds On The Error Incurred From Rabi Oscillations \label{Appendix C}}
Here we present a proof for Theorem \ref{Theorem 2}:
\begin{proof}
After a $\frac{\pi}{2}$ rotation around the $y$-axis, we'll get:
\begin{equation}
H_{r}=-\frac{\Delta}{2}\mathbb{X}-\lambda\cos(\omega t)\mathbb{Z}\ .
\end{equation}
As it is explained in \cite{Deng2016}, an approximate solution to
this Hamiltonian, which is valid for both weak and strong coupling
is given by:
\begin{equation}
|\psi\left(t\right)\rangle_{r}=\alpha_{0}e^{-i\epsilon_{0}t}|u_{0}\left(t\right)\rangle_{r}+\alpha_{1}e^{-i\epsilon_{1}t}|u_{1}\left(t\right)\rangle_{r}\ ,
\end{equation}
where:
\begin{equation}
\begin{cases}
\epsilon_{0} & =\frac{1}{2}\left(-\omega-\sqrt{\left[\omega-\Delta J_{0}\left(\frac{2\lambda}{\omega}\right)\right]^{2}+\Delta^{2}J_{1}^{2}\left(\frac{2\lambda}{\omega}\right)}\right) \vspace{5pt}\\
\epsilon_{1} & =\frac{1}{2}\left(-\omega+\sqrt{\left[\omega-\Delta J_{0}\left(\frac{2\lambda}{\omega}\right)\right]^{2}+\Delta^{2}J_{1}^{2}\left(\frac{2\lambda}{\omega}\right)}\right)
\end{cases}
\end{equation}
and
\begin{equation}
|u_{j}\left(t\right)\rangle_{r}=\sum_{n=-\infty}^{\infty}e^{in\omega t}|u_{j,n}\rangle_{r}\qquad\text{for }j=0,1\ ,
\end{equation}
where $J_0$ and $J_1$ are Bessel functions of the first kind and:
\begin{equation}
\begin{cases}
|u_{0,n}\rangle_{r} & =\frac{1}{\sqrt{2}}\begin{pmatrix}\cos\left(\frac{\theta}{2}\right)J_{n+1}\left(\frac{\lambda}{\omega}\right)+\sin\left(\frac{\theta}{2}\right)J_{n}\left(\frac{\lambda}{\omega}\right)\\
-\cos\left(\frac{\theta}{2}\right)J_{n+1}\left(-\frac{\lambda}{\omega}\right)+\sin\left(\frac{\theta}{2}\right)J_{n}\left(-\frac{\lambda}{\omega}\right)
\end{pmatrix} \vspace{5pt}\\
|u_{1,n}\rangle_{r} & =\frac{1}{\sqrt{2}}\begin{pmatrix}-\sin\left(\frac{\theta}{2}\right)J_{n+1}\left(\frac{\lambda}{\omega}\right)+\cos\left(\frac{\theta}{2}\right)J_{n}\left(\frac{\lambda}{\omega}\right)\\
\sin\left(\frac{\theta}{2}\right)J_{n+1}\left(-\frac{\lambda}{\omega}\right)+\cos\left(\frac{\theta}{2}\right)J_{n}\left(-\frac{\lambda}{\omega}\right)
\end{pmatrix}
\end{cases}\ ,
\end{equation}
where:
\begin{equation}
\tan\left(\theta\right)=\frac{\Delta J_{1}\left(\frac{2\lambda}{\omega}\right)}{\omega-\Delta J_{0}\left(\frac{2\lambda}{\omega}\right)}\ .
\end{equation}

For the rest of this section, we assume exact resonance conditions,
\emph{i.e.}\ $\Delta=\omega$:
\begin{equation}
\tan\left(\theta\right)=\frac{J_{1}\left(\frac{2\lambda}{\omega}\right)}{1-J_{0}\left(\frac{2\lambda}{\omega}\right)}\ .
\end{equation}
Because we start in the ground state of our non-rotated Hamiltonian,
Eq.\ (\ref{eq:nonrotated hamiltonian}), we need to rotate our system
again to get:
\begin{equation}
\begin{cases}
|u_{0,n}\rangle & =\frac{1}{\sqrt{2}}\begin{pmatrix}1 & 1\\
-1 & 1
\end{pmatrix}|u_{0,n}\rangle_{r} \vspace{4pt}\\
|u_{1,n}\rangle & =\frac{1}{\sqrt{2}}\begin{pmatrix}1 & 1\\
-1 & 1
\end{pmatrix}|u_{1,n}\rangle_{r}
\end{cases}\ .
\end{equation}
The time dependent state of the system is:
\begin{equation}
|\psi\left(t\right)\rangle=\beta_{0}e^{-i\epsilon_{0}t}\sum_{n=-\infty}^{\infty}e^{in\omega t}|u_{0,n}\rangle+\beta_{1}e^{-i\epsilon_{1}t}\sum_{n=-\infty}^{\infty}e^{in\omega t}|u_{1,n}\rangle\ .
\end{equation}
After setting the initial condition to:
\begin{equation}
|\psi\left(0\right)\rangle=\begin{pmatrix}1\\
0
\end{pmatrix}\ ,
\end{equation}
and using the Bessel function identity: 
\begin{equation}
\sum_{n=-\infty}^{\infty}J_{n}\left(x\right)=1\ ,
\end{equation}
we'll find the coefficients $\beta_{0}$ and $\beta_{1}$ to be:
\begin{equation}
\begin{cases}
\beta_{0} & =\sin\left(\frac{\theta}{2}\right)\\
\beta_{1} & =\cos\left(\frac{\theta}{2}\right)
\end{cases}
\end{equation}

After a bit of simplification one can write the state of the system
explicitly as:
\begin{equation}
\begin{array}{l}
|\psi\left(t\right)\rangle\\
=\begin{pmatrix}\sum_{k=-\infty}^{\infty}e^{2ik\omega t}\left(\frac{\sin\theta}{2}J_{2k+1}\left(\frac{\lambda}{\omega}\right)\left(e^{-i\epsilon_{0}t}-e^{-i\epsilon_{1}t}\right)+J_{2k}\left(\frac{\lambda}{\omega}\right)\left(\sin^{2}\left(\frac{\theta}{2}\right)e^{-i\epsilon_{0}t}+\cos^{2}\left(\frac{\theta}{2}\right)e^{-i\epsilon_{1}t}\right)\right) \vspace{4pt}\\
\sum_{k=-\infty}^{\infty}e^{\left(2k-1\right)i\omega t}\left(\frac{\sin\theta}{2}J_{2k}\left(\frac{\lambda}{\omega}\right)\left(e^{-i\epsilon_{1}t}-e^{-i\epsilon_{0}t}\right)-J_{2k-1}\left(\frac{\lambda}{\omega}\right)\left(\sin^{2}\left(\frac{\theta}{2}\right)e^{-i\epsilon_{0}t}+\cos^{2}\left(\frac{\theta}{2}\right)e^{-i\epsilon_{1}t}\right)\right)
\end{pmatrix}\ .
\end{array} \label{eq:psi-with-summation}
\end{equation}
From the ordinary RWA, we expect the maximum transition to the excited
state to happen around $t\simeq\frac{\pi}{\lambda}$. 

Using Jacobi-Anger expansion, one can prove:
\begin{equation}
\begin{cases}
\sum_{n=-\infty}^{\infty}e^{\left(2n\right)i\phi}J_{2n}\left(x\right) & =\cos\left(x\sin\left(\phi\right)\right) \vspace{4pt}\\
\sum_{n=-\infty}^{\infty}e^{\left(2n-1\right)i\phi}J_{2n-1}\left(x\right) & =i\sin\left(x\sin\left(\phi\right)\right)
\end{cases}\ .
\end{equation}
We can use these to take care of the summations in Eq.\ (\ref{eq:psi-with-summation}):
\begin{equation}
\begin{array}{rl}
\left|\left\langle 1|\psi\left(\frac{\pi}{\lambda}\right)\right\rangle \right| & =\left|\frac{\sin\theta}{2}e^{\nicefrac{-i\pi\omega}{\lambda}}\left(e^{\nicefrac{-i\pi\epsilon_{1}}{\lambda}}-e^{\nicefrac{-i\pi\epsilon_{0}}{\lambda}}\right)\cos\left(\frac{\lambda}{\omega}\sin\left(\frac{\pi\omega}{\lambda}\right)\right)\right. \vspace{4pt}\\
 & \left.-i\left(\sin^{2}\left(\frac{\theta}{2}\right)e^{\nicefrac{-i\pi\epsilon_{0}}{\lambda}}+\cos^{2}\left(\frac{\theta}{2}\right)e^{\nicefrac{-i\pi\epsilon_{1}}{\lambda}}\right)\sin\left(\frac{\lambda}{\omega}\sin\left(\frac{\pi\omega}{\lambda}\right)\right)\right|\ .
\end{array}
\end{equation}
In order to be able to express the last equation as a series in powers
of $\left(\frac{\lambda}{\omega}\right)$, let's define $\kappa\equiv\sin\left(\frac{\pi\omega}{\lambda}\right)$.
$\kappa$ is highly sensitive to the ratio $\frac{\omega}{\lambda}$,
but its value is bounded between:
\begin{equation}
-1\le\kappa\le1
\end{equation}
Up to the first nonzero order we get:
\begin{equation}
\left|\Braket{ 1|\psi\left(\frac{\pi}{\lambda}\right)} \right|\sim\left|1-\left(\frac{1+\kappa^{2}}{2}+i\kappa e^{\nicefrac{i\pi\omega}{\lambda}}\right)\left(\frac{\lambda}{\omega}\right)^{2}+O\left(\left(\frac{\lambda}{\omega}\right)^{4}\right)\right|\ ,
\end{equation}
\begin{equation}
\Rightarrow1-\left|\Braket{ 1|\psi\left(\frac{\pi}{\lambda}\right)} \right|\sim\left|\left(\frac{\cos^{2}\left(\frac{\pi\omega}{\lambda}\right)}{2}+i\sin\left(\frac{\pi\omega}{\lambda}\right)\cos\left(\frac{\pi\omega}{\lambda}\right)\right)\right|\left(\frac{\lambda}{\omega}\right)^{2}+O\left(\left(\frac{\lambda}{\omega}\right)^{4}\right)\ .
\end{equation}
Maximizing the r.h.s.\ expression over its valid range of arguments
and ignoring the higher order terms, one finds:
\begin{equation}
1-\left|\Braket {1|\psi\left(\frac{\pi}{\lambda}\right)} \right|\le\frac{1}{\sqrt{3}}\left(\frac{\lambda}{\omega}\right)^{2}\ .
\end{equation}
This is good enough for our purposes, but if one can do the experiment
with high precision and high control over $\omega$ and $\lambda$;
by choosing $\omega=\left(n+\frac{1}{2}\right)\lambda$, this can
in principle be improved to:
\begin{equation}
\frac{\pi}{48}\left(\frac{\lambda}{\omega}\right)^{4}+O\left(\left(\frac{\lambda}{\omega}\right)^{5}\right)\le1-\left|\left\langle 1|\psi\left(\frac{\pi}{\lambda}\right)\right\rangle \right|\le\frac{1}{\sqrt{3}}\left(\frac{\lambda}{\omega}\right)^{2}+O\left(\left(\frac{\lambda}{\omega}\right)^{4}\right)\ .
\end{equation}
\end{proof}

\part*{Bibliography}

\bibliographystyle{plain}
\bibliography{My_Collection}

\end{document}